\documentclass[a4paper]{spie}  

\usepackage{amsmath,amsfonts,amssymb}
\usepackage[colorlinks=true, allcolors=blue]{hyperref}

\usepackage[english]{babel}
\usepackage[utf8]{inputenc}
\usepackage[]{graphicx}
\usepackage[expansion=false]{microtype}
\usepackage{gensymb}
\usepackage{epsfig}

\usepackage{float}

\usepackage{gensymb}
\usepackage{todonotes}
\usepackage{boxedminipage2e}
\usepackage{siunitx}
\usepackage{booktabs}
\usepackage{wasysym}
\usepackage{subcaption}
\usepackage{todonotes}

\usepackage{natbib}

\usepackage{tikz}
\usetikzlibrary{calc}
\usetikzlibrary{arrows}
\usetikzlibrary{arrows.meta}
\tikzstyle{unit}=[draw=blue,text=blue,fill=white, very thick,shape=circle,minimum size=0.45cm]
\tikzstyle{unit2}=[draw=red!30,text=red!30,fill=white, very thick,shape=circle,minimum size=0.45cm]
\tikzstyle{unit3}=[draw=red!30,text=red!30,fill=white, very thick,shape=circle,minimum size=0.85cm]

\tikzstyle{Labels}=[rectangle, draw=blue, rounded corners=0.3cm, fill=blue!10, minimum height=0.6cm, minimum width=1.5cm, text=blue, align=center, thick]

\tikzstyle{LabelsLong}=[rectangle, draw=blue, rounded corners=0.3cm, fill=blue!10, minimum height=0.6cm, text width=4cm, text=blue, align=center, thick]

\tikzstyle{LabelsSub}=[rectangle, draw=blue, rounded corners=0.2cm, fill=blue!10, minimum height=0.4cm, text=blue, align=center, thick, font=\small]

\tikzstyle{ARROW}=[->, >=stealth, very thick, draw=blue]

\tikzstyle{marker}=[color=blue]
\tikzstyle{line}=[color=blue, very thick]

\title{MOSES - The MONET Star and Exoplanet Spectrograph }

\author[a]{S. Schäfer}
\author[a]{L. Schmidt}
\author[a]{H. Anwand-Heerwart}
\author[b]{D. Jones}
\author[a]{A. Reiners}

\affil[a]{Georg-August Universität Göttingen, Institut für Astrophysik und Geophysik, Friedrich-Hund-Platz 1, 37077 Göttingen, Germany}
\affil[b]{Prime Optics, 17 Crescent Road, Eumundi Q 4562 Australia}
\authorinfo{Further author information: (Send correspondence to S. Schäfer)\\ E-mail: sebastian.schaefer@phys.uni-goettingen.de}

\begin{document} 

\def\aj{AJ}%
\def\actaa{Acta Astron.}%
\def\araa{ARA\&A}%
\def\apj{ApJ}%
\def\apjl{ApJ}%
\def\apjs{ApJS}%
\def\ao{Appl.~Opt.}%
\def\apss{Ap\&SS}%
\def\aap{A\&A}%
\def\aapr{A\&A~Rev.}%
\def\aaps{A\&AS}%
\def\azh{AZh}%
\def\baas{BAAS}%
\def\bac{Bull. astr. Inst. Czechosl.}%
\def\caa{Chinese Astron. Astrophys.}%
\def\cjaa{Chinese J. Astron. Astrophys.}%
\def\icarus{Icarus}%
\def\jcap{J. Cosmology Astropart. Phys.}%
\def\jrasc{JRASC}%
\def\mnras{MNRAS}%
\def\memras{MmRAS}%
\def\na{New A}%
\def\nar{New A Rev.}%
\def\pasa{PASA}%
\def\pra{Phys.~Rev.~A}%
\def\prb{Phys.~Rev.~B}%
\def\prc{Phys.~Rev.~C}%
\def\prd{Phys.~Rev.~D}%
\def\pre{Phys.~Rev.~E}%
\def\prl{Phys.~Rev.~Lett.}%
\def\pasp{PASP}%
\def\pasj{PASJ}%
\def\qjras{QJRAS}%
\def\rmxaa{Rev. Mexicana Astron. Astrofis.}%
\def\skytel{S\&T}%
\def\solphys{Sol.~Phys.}%
\def\sovast{Soviet~Ast.}%
\def\ssr{Space~Sci.~Rev.}%
\def\zap{ZAp}%
\def\nat{Nature}%
\def\iaucirc{IAU~Circ.}%
\def\aplett{Astrophys.~Lett.}%
\def\apspr{Astrophys.~Space~Phys.~Res.}%
\def\bain{Bull.~Astron.~Inst.~Netherlands}%
\def\fcp{Fund.~Cosmic~Phys.}%
\def\gca{Geochim.~Cosmochim.~Acta}%
\def\grl{Geophys.~Res.~Lett.}%
\def\jcp{J.~Chem.~Phys.}%
\def\jgr{J.~Geophys.~Res.}%
\def\jqsrt{J.~Quant.~Spec.~Radiat.~Transf.}%
\def\memsai{Mem.~Soc.~Astron.~Italiana}%
\def\nphysa{Nucl.~Phys.~A}%
\def\physrep{Phys.~Rep.}%
\def\physscr{Phys.~Scr}%
\def\planss{Planet.~Space~Sci.}%
\def\procspie{Proc.~SPIE}%
\def\maps{M\&PS}%
\let\astap=\aap
\let\apjlett=\apjl
\let\apjsupp=\apjs
\let\applopt=\ao

\maketitle

\begin{abstract}
We introduce MOSES, the new High-Resolution Echelle Spectrograph designated for the 1.2m MONET telescope at McDonald Observatory, Texas, USA. The science drivers are radial velocity experiments and activity monitoring in Sun-like stars. Set for installation in the final quarter of 2026, MOSES features a white pupil design and aims for a spectral resolution greater than 80,000 over the 380-680 nm wavelength range. It incorporates a pixel sampling rate of 3.5 and uses two fibers to facilitate a simultaneous calibration mode. Encased within a vacuum vessel and operating in a temperature-stabilized environment, MOSES is expected to achieve a radial velocity precision below 2 m/s, aided by a Fabry-Pérot etalon calibration system. This paper outlines the implementation of the fiber injection unit, the optical layout of the spectrograph, and the present status of the various subsystems under development. 
\end{abstract}


\section{Science case and project goals}
\label{sec:intro}

The search for extrasolar planets has revealed thousands of planetary systems around very different stars. Radial velocity measurements are required to determine the mass of the planets, for which current and future programs reach Doppler precisions at the 1\,m\,s$^{-1}$ level and below. 
In order to characterize the planetary population and to understand planet
formation, in particular in comparison to the solar system, we need to search for planets at large distances to their host
stars. In particular, we need to understand whether the solar system with its
two giant planets, Jupiter and Saturn, on orbits beyond 5\,AU is the rule or
an exception. At these distances, the RV method is by far the most efficient
method for nearby stars because transit probabilities are extremely low. A robust result about the occurrence of planets similar to those in the solar system is the frequency of cool giants around Sun-like stars: the average number of cool giant planets with masses in the range 30--6000\,M$_{\oplus}$, at orbital distances $a > 2$\,AU around stars in the mass range 0.5--1.5\,M$_{\odot}$ is $0.23 \pm 0.02$, i.e., approximately every fourth Sun-like stars hosts a Saturn- or Jupiter-like cool giant planet \citep{2021ApJS..255...14F}. These are the exoplanet systems that are potentially very similar to the solar system, many of them may host additional planets that have so far remained undiscovered. 

Only a relatively small number of projects are
systematically studying the presence of distant planets on long orbits, among
them are the CORALIE survey for southern extrasolar planets
\citep[e.g.,][]{2010A&A...511A..45S}, the McDonald observatory planet search
\citep[e.g.,][]{2016ApJ...818...34E}, and projects combining dedicated new
observations with archive or published data from different sources
\citep[e.g.,][]{2016Natur.536..437A, 2018Natur.563..365R,
  2019AJ....158..181B}. This is the type of study that requires dedicated facilities with continuous access and is therefore suited for smaller telescopes. In \citet{2020ApJS..247...11R}, we carried out a detailed
analysis of the achievable RV photon noise limit for different instruments and
a large number of nearby stars. With the goal to estimate the number of stars
available for our projects, we performed the simulation for spectrographs at small telescopes: there are
more than a thousand stars in which the precision limit of 3\,m\,s$^{-1}$ can
be reached within a 10\,min exposure with a high-resolution spectrograph at a 1.2\,m telescope, which underlines
the great potential of small telescopes. 

\begin{figure}[b]
    \centering
    \includegraphics[width = 1.0\textwidth]{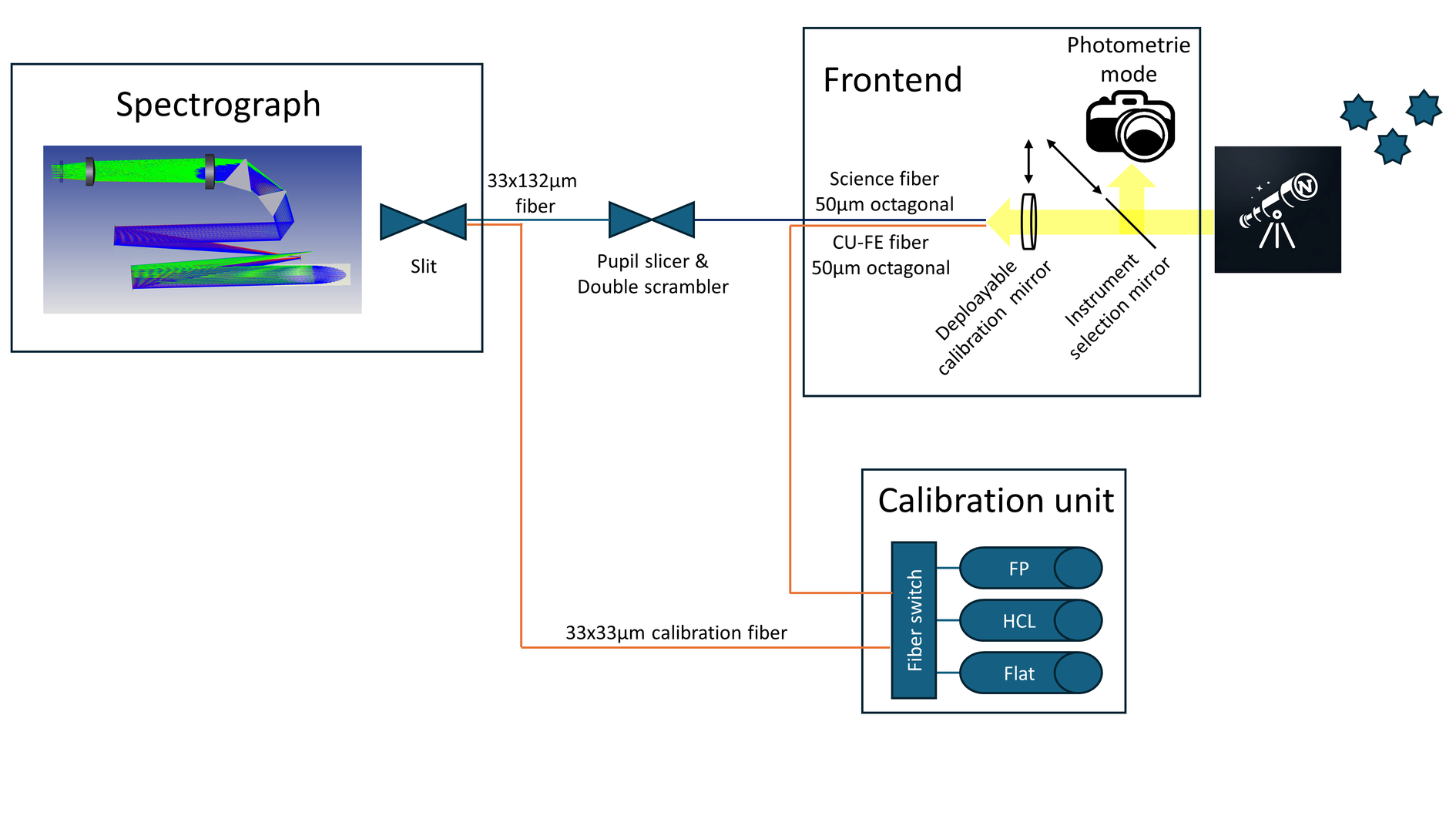}
    \caption{Schematic overview of the instrument design.}
    \label{fig:Setup}
\end{figure}

MOSES is a high-resolution spectrograph designed for our 1.2\,m telescope located at McDonald observatory, Texas, USA. Our main scientific program
is a long-term RV study of several hundred stars at a precision of
3\,m\,s$^{-1}$. This precision approximately equals the RV amplitude of a
planet with a projected mass ($M\,\sin{i}$) of about two Neptune masses at
1\,AU (Earth orbit, 1\,a period), about 0.25 Jupiter masses at 5\,AU
(Jupiter's orbit, 12\,a period), and about one Saturn mass at 9.5\,AU
(Saturn's orbit, 29\,a period). Thus, planets like the solar system giant
planets Jupiter and Saturn can be found out to the distance of Saturn in a
long-term survey with telescopes as small as 1.2\,m. To achieve this precision on sky, we designed the spectrograph with a resolving power
of $>$\,82,000 and a pixel sampling rate of 3.5 per resolution element. Our approach follows the established strategy of simultaneous calibration using Fabry-Pérot etalons for drift correction and hollow cathode lamps to obtain the wavelength solution using a second fiber in parallel to the stellar light (e.g. \cite{2003Msngr.114...20M}, \cite{2016SPIE.9908E..12Q} and \cite{Pepe_2021}). The wavelength coverage spans from 380\,-\,680\,nm. Additionally, a combined pupil slicer and double scrambler, following the design of \cite{Beckert2022}, will be utilized, as illustrated in Figure\,\ref{fig:Setup}.

Stellar activity plays a crucial role in the evolution of Sun-like and
low-mass stars \citep[e.g.,][]{1990ApJS...74..501P, 1997A&A...326.1023B,
  2012ApJ...746...43R}. In the context of extrasolar planets, activity is
relevant because the variability of features associated with stellar activity
interferes with the signatures used for the detection and characterization of
planets, most important the variability of photometric light curves for
transit detections and the variability of spectral line positions for the
Doppler method \citep[e.g.,][]{2010ApJ...710..432R, 2014MNRAS.443.2517H}. 
Long-term magnetic cycles are observed in many stars other than the Sun
\citep{1985ARA&A..23..379B, 2018A&A...616A.108B}. The primary indicator for
activity in sun-like stars is chromospheric emission in the Ca\,\textsc{ii}
H\,\&\,K lines at 396.8 and 393.4\,nm respectively; this indicator will also be obtained with MOSES. Thus,
for each of our observations of long-term Doppler variability, we will
simultaneously gather a measurement of stellar activity, and for our
monitoring campaign the data will be available on timescales of the solar
magnetic cycle. Furthermore, the long wavelength coverage and high Doppler
precision will allow using RV measurements together with their wavelength
dependence, or the chromatic index \citep{2018A&A...609A..12Z}, to
discriminate between RV signatures caused by Keplerian motion from those
caused by stellar variability.

\section{MONET/North telescope site}
\label{sec:monet}
The MONET/North telescope, part of the Institut für Astrophysik und Geophysik Göttingen (IAG), was established at the McDonald Observatory in Texas, USA, in 2005. Its twin telescope, MONET/South, also owned and operated by IAG, has been functioning in Sutherland, South Africa since 2008. Both telescopes have been used exclusively for photometric observations, however since 2023 MONET/S also features a newly installed small low resolution spectrograph (see \cite{Meerwart2024} for details).

\subsection{Telescope}

The telescope features a 1.2\,m main mirror and was constructed by Halfmann Teleskoptechnik using an altitude-azimuth mount. For a complete list of parameters, see Table\,\ref{tab:mcdonald}, and for details on the initial construction, refer to \cite{Hessman2004}.

\begin{table}[h]
\centering
\begin{tabular}{|l|c|}
\hline
Site altitude & 2077\,m \\ \hline
Site atmospheric pressure & 0.799\,mbar \\ \hline
Primary Diameter & 1.2\,m \\ \hline
Secondary diamter & 0.5\,m \\ \hline
Telescope focal length                                  & 8400 mm                   \\ \hline
Focal ratio & 7 \\ \hline
Expected 80 percentile seeing & 2.2 arcsecs \\ \hline
Image diameter on focal plane & 90 µm \\ \hline

\end{tabular}
\caption{Telescope parameters}
\label{tab:mcdonald}
\end{table}

The MONET telescopes are part of a family of very similarly designed telescopes, including the TIGRE telescope in Mexico, operated by the University of Hamburg, and the two STELLA telescopes in Tenerife, managed by the Leibniz Institute for Astrophysics Potsdam. Both the MONET and STELLA telescopes are undergoing upgrades to their electronics with Beckhoff components to enhance reliability, ensure better availability of spare parts, and replace the current black-box systems (see \cite{2022SPIE12189E..0FW}).

Historically, the MONET telescopes have been dedicated to a single instrument each (in both cases photometric cameras). In contrast, one of the STELLA telescopes boasts a 180\,\textdegree\ rotatable M3 mirror that allows access to the second Nasmyth port for additional instrumentation. This capability will soon be extended to the MONET telescope through a new frontend, as introduced in Section\,\ref{sec:frontend}, enabling a fiber-fed spectroscopy mode in addition to photometry. Given that both MONET telescopes are equipped with the same flange, instruments can theoretically be exchanged between them at any time.

\subsection{Telescope building}
The buildings housing the MONET telescopes at both observatories share a uniform design, developed modularly by a South African company. After production, the structure for MONET/N was shipped to the US and assembled on site. Each telescope is mounted on an insulated foundation pier within the building and elevated to the first floor. A grating platform encircles the telescope, facilitating maintenance work. The basement provides ample space for electronics, an uninterruptible power supply (UPS), and workspace.

\begin{figure}[!h]
    \centering
    \includegraphics[width = 1.0 \textwidth]{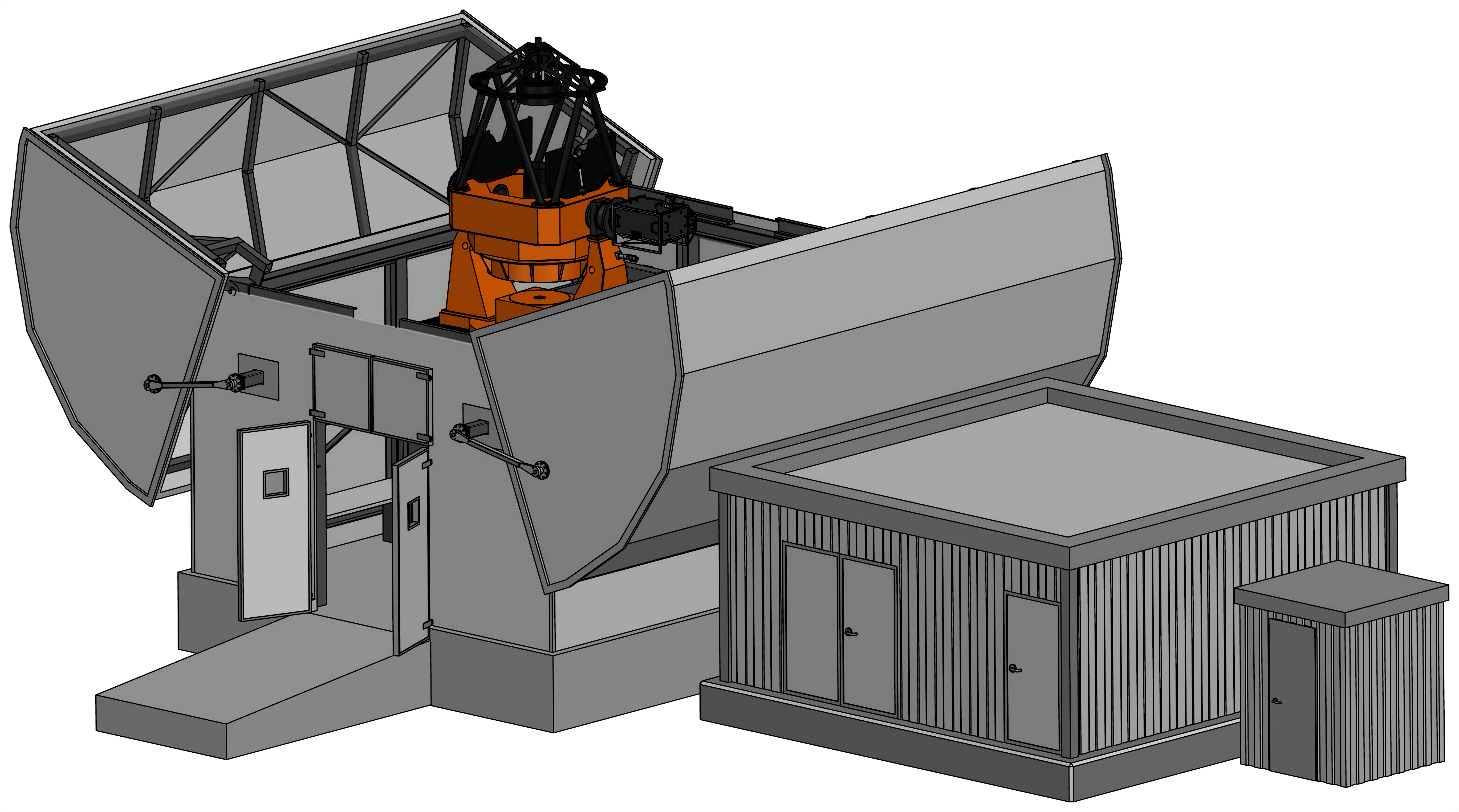} 
    \caption{The MONET/N telescope next to the planned MOSES building and the existing transformer station.}
    \label{fig:McDonald}
\end{figure}

A distinctive feature of these buildings is their clam-shell roof (see Figure\,\ref{fig:McDonald}). This unconventional design presents both benefits and challenges. Although more susceptible to disturbances from relatively calm winds compared to traditional dome-enclosed telescopes, the clam-shell roof does not restrict the telescope's operational speed for rapid targeting, an advantage in scenarios requiring swift target switching. Moreover, the clam-shell design proves more cost-effective than dome solutions, which are typically more expensive.

\subsection{MOSES Building}

The new building for the MOSES spectrograph will be situated adjacent to the telescope dome, as illustrated in Fig.\,\ref{fig:McDonald}. It will adopt the versatile design of a mobile container laboratory, which can be disassembled and transported as needed. This modular approach provides several benefits, such as enabling a standardized design for similar instruments potentially at MONET/S in South Africa, facilitating the reuse of the design across different observatories worldwide for various projects, and crucially, allowing the laboratory to be transported to Göttingen for extensive testing in our integration hall prior to its final deployment at the McDonald Observatory.

\begin{figure}
    \centering
    \includegraphics[width=0.85\textwidth]{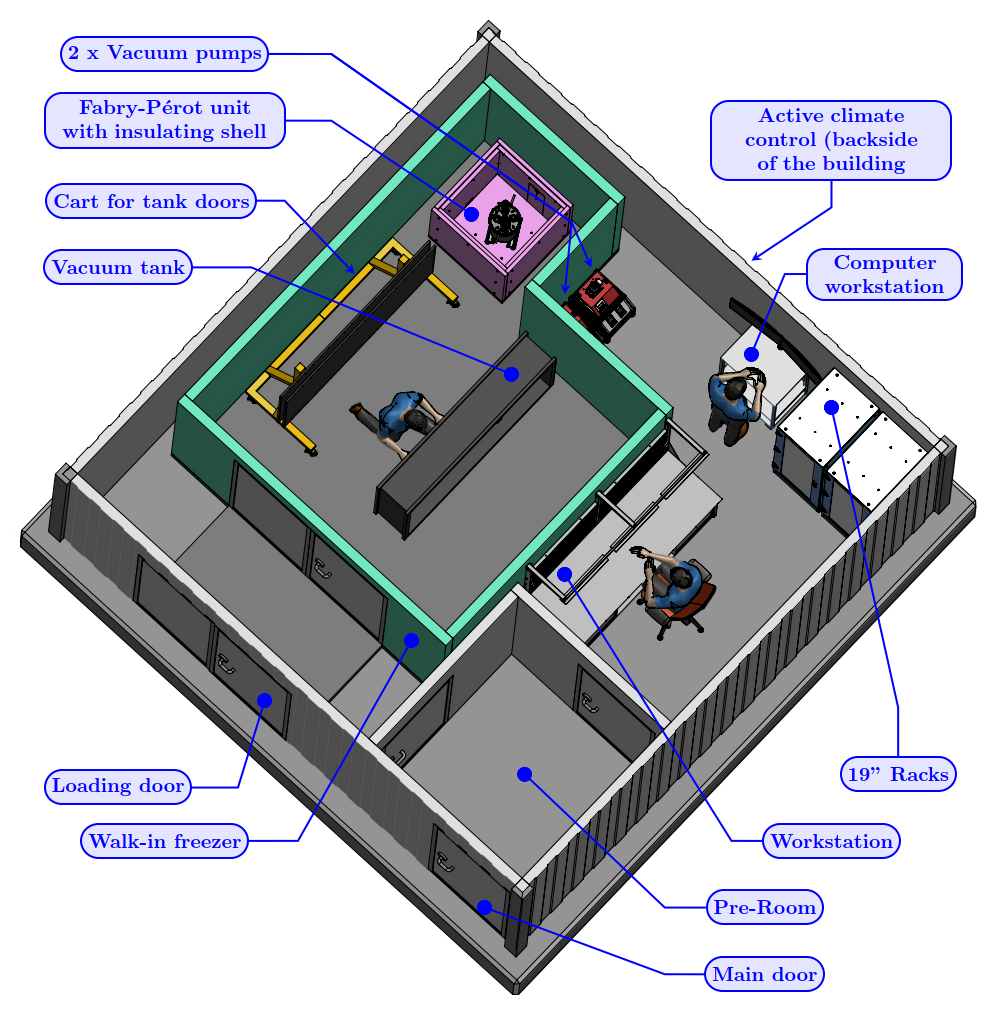}
    \caption{The planned MOSES building: The instrument is located inside a commercial walk-in freezer (green walls) unit with the Fabry-Pérot unit also located inside. The Vacuum pumps and all electronics are outside this enclosure in the working area of the building, as is a workbench and the active climate controll unit.}
    \label{fig:moses_lab} 
\end{figure}

The building will incorporate an active climate control system, designed to maintain a stable interior temperature of 20\,$\pm$\,1\textdegree\,C over a 24-hour period, with a goal of achieving $\pm$\,0.5\textdegree\,C stability over the course of a year. It consists of three rooms: the main instrument will be housed within a commercial walk-in freezer unit (highlighted by green walls in  Figure\,\ref{fig:moses_lab})  which adds a layer of passive insulation. Additionally, the vacuum vessel of the spectrograph will be further insulated with Styrodur, aiming for a temperature stability of about  $\pm$0.1\,\textdegree\,C for the instrument itself.

\clearpage
\section{Frontend}
\label{sec:frontend}

The frontend unit (FE) design is driven by top-level requirements for throughput and image stabilization in the spectroscopy mode, and it also has to support the photometric mode. The latter is achieved by moving a flat 45-degree mirror into the telescope beam, directing it towards a Kepler KL4040 FI CMOS camera, equipped with a FLI CL-1-14 filter wheel. The camera is a front illuminated CMOS with 4096 x 4096 pixel, a pixel size of 9 $\mu$m, a typical readout noise of 3.7e- and a dynamic range of 85.2\, db. It is capable of a 23\,fps frame rate and a typical dark current of 0.4\,eps at -10\textdegree C.

For the spectroscopy mode the FE has to convert the f/7 beam of MONET/N to f/3.9 to allow efficient fiber coupling while keeping the focal ratio degradation (FRD) at a minimum. Additionally, the FE has to provide access to the telescopes pupil for the implementation of atmospheric dispersion correction (ADC), a fast tip/tilt mirror (TTM) and a beamsplitter to direct about 2\% of the light into a dedicated guiding camera. 

The FE will be mounted to one of the two Nasmyth ports of the MONET/N telescope with a new cable wrap unit also being attached directly between the de-rotator and the FE unit.

\subsection{Spectroscopy mode requirements and optics}
\begin{figure}[!h]
    \centering
        \begin{subfigure}[t]{.45\textwidth} 
            \centering
            \input{pictures/sciencemode_new.tex}
            \caption{Science mode: Light form the telescope passes the telescope focus, is collimated, passes the ADC, the TTM, the beamsplitter and is focused onto the science fiber.}
            \label{fig:FE_science}
        \end{subfigure}
        \quad \quad
        \begin{subfigure}[t]{0.45\textwidth}
            \centering
            \input{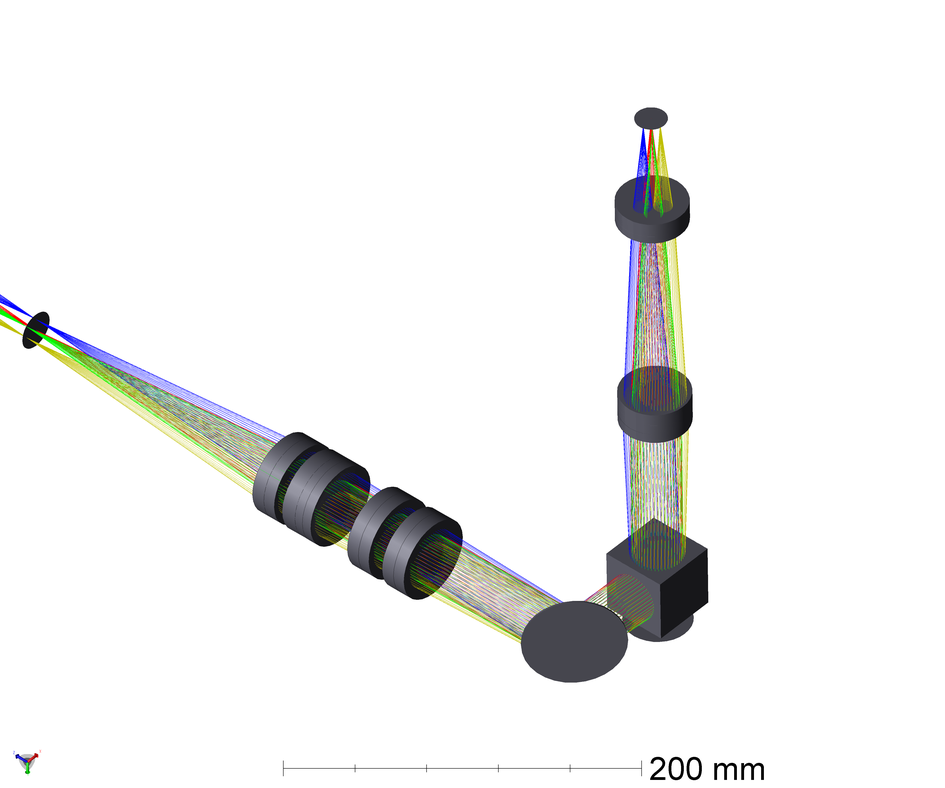}

            \caption{Guiding during science observations:  Light form the telescope passes the telescope focus, is collimated, passes the ADC and the TTM. About 2\,\% of the light are deflected in the beamsplitter and is focused into the A\&G camera.}
            \label{fig:FE_guide}
        \end{subfigure}
        
        \begin{subfigure}[t]{.45\textwidth}
            \centering
            \input{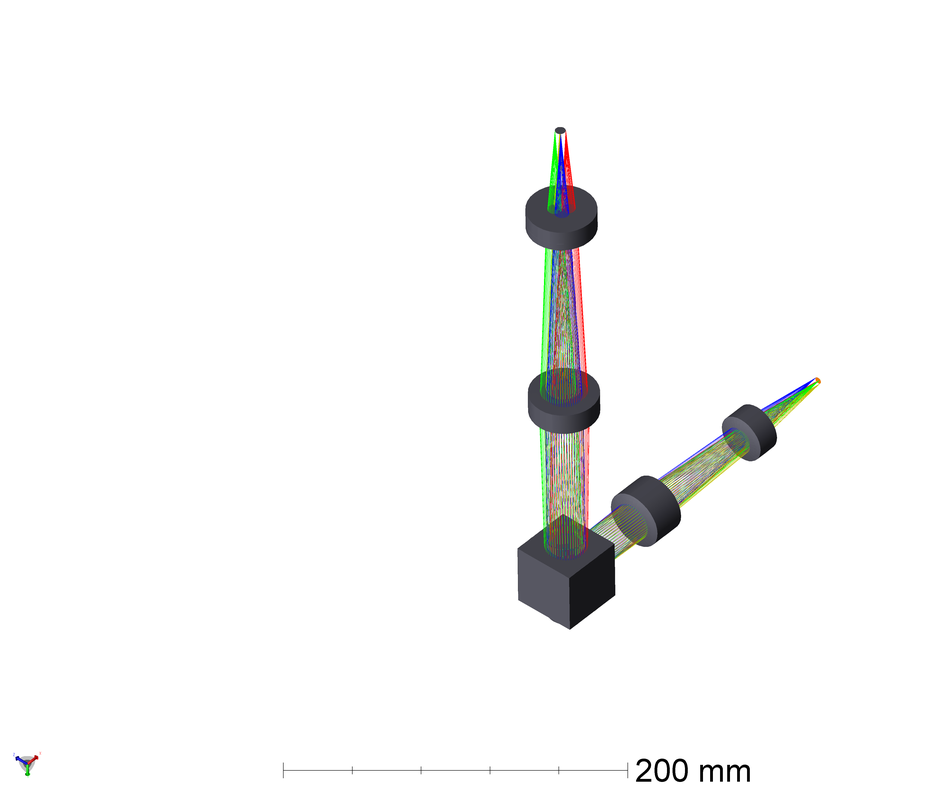}

            \caption{Aiming mode: Light from back illuminated single mode fibers is collimated and send through the beamsplitter. 2\% are deflected 'down' towards a retro reflector sending the light again to the beamsplitter. This time 98\% are transmitting towards the A\&G camera.}
            \label{fig:FE_aim}
        \end{subfigure}
        \quad \quad
        \begin{subfigure}[t]{0.45\textwidth}
            \centering
            \input{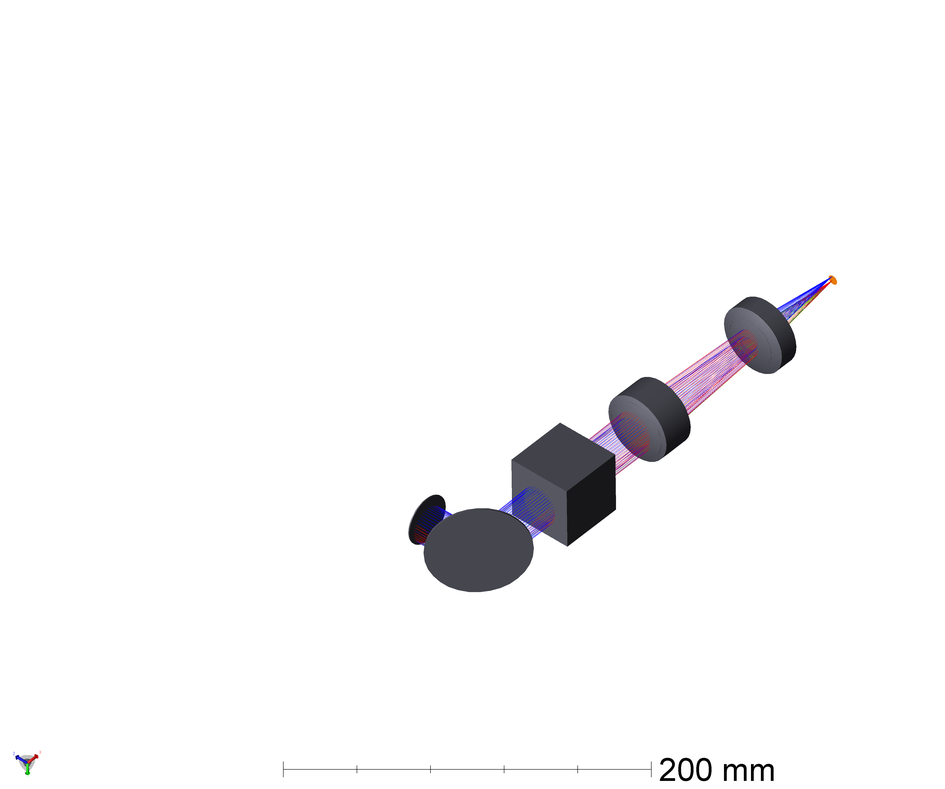}

            \caption{Calibration light mode: Light from the calibration unit is send through the calibration FE fiber. It is collimated and passes through the beamsplitter and the TTM and hits a deployable retro reflector. Using the TTM the light is send the same way back onto the science fiber, using the A\&G camera for verification (not shown).}
            \label{fig:FE_calibration}
        \end{subfigure}
        \vspace{0.25cm}
        \caption{Modes of operation of the FE: Optical setup between telescope focus (lower left), ADC, tip-tilt foldmirror (TTM), beamsplitter, fiber injection (top right) and guiding camera (top center)}
    \end{figure}

There are three operational modes for the spectroscopy mode:

In \textbf{normal science observation mode} (see Figs.\,\ref{fig:FE_science}\,\&\,\ref{fig:FE_guide}), the light from the telescope passes through the telescope's focus. Subsequently, a collimator forms a collimated beam space followed by reimaging optics for fiber injection and a guidance camera. In practice, a collimator of focal length 200\,mm is used to create an image of the telescope pupil at a useful distance from its last lens. For the ADC two counter-rotating Risley Prisms (or Direct Vision Doublet Prisms – DVDPs) are positioned close to the last lens of the collimator. Following the ADC is space for the deployable fixed calibration retro mirror (CRM), the TTM and a cube beamsplitter respectively.  The cube beamsplitter directs roughly 2\% of the incoming light to an Acquisition and Guidance (A\&G) camera. 

Thus, we are therefore not guiding on the wings of the PSF but on the core of the PSF, which is directly imaged onto the guiding camera. The collimator and A\&G camera must accommodate a square field of view of 400 arcsecs on a side. This corresponds to a square field of view on the telescope image surface with a side dimension of 16.3 mm. Light undeflected by the beamsplitter is imaged by the fiber imager on to a “fiber block”, which is an array of fibers on the fiber imager image surface. The fiber array, or block, includes 4 guide fibers, the science fiber, and a calibration fiber arranged symmetrically around the science fiber (see Table\,\ref{tab:fiber}). The A\&G camera is used to align the star on top of the science fiber (see below). 

\vspace{1cm}
\begin{table}[h]
\centering
\begin{tabular}{|c|c|}
\hline
\textbf{Fiber} & \textbf{Location (x, y) in mm} \\ \hline
Object/Science & (0, 0) \\ \hline
Guide & (1.5, 1.5) \\ \hline
 & (-1.5, 1.5) \\ \hline
 & (-1.5, -1.5) \\ \hline
 & (1.5, -1.5) \\ \hline
Calibration & (0, 3) \\ \hline
\end{tabular}
\vspace{0.5cm}
\caption{Fiber block geometry and location of all the fibers. }
\label{tab:fiber}
\end{table}

\clearpage
\clearpage

The four guide fibers are single mode fibers used to determine the pixel position of the science fiber on the guiding camera. \textbf{In aiming mode} (see Fig.\,\ref{fig:FE_aim}) they are illuminated by a NIR LED outside the spectrographs waveband (900\,nm). Light exiting the four fibers is collimated by the fiber imager optics and hits the beamsplitter cube. 2\% are reflected to an external retro mirror and when hitting the beamsplitter again 98\% are transmitted towards the A\&G camera where the fibers are reimaged. Since they form a square, the central position of the science fiber in pixel space on the camera can be determined, which is needed for the normal guiding.   

The third mode is called \textbf{calibration mode} (see Fig.\,\ref{fig:FE_calibration}). It can be used to send the light from the calibration unit (which is normally fed directly into the spectrograph by a direct calibration fiber) through the FE and through the science fiber. This enables calibration light to be sent through the same path as the starlight and is used for the wavelength calibration of the spectrograph. The light is sent from the calibration unit towards the FE through the FE calibration fiber, is collimated by the fiber imager, sent through the beamsplitter and the TTM before hitting the CRM. The TTM can now be used to steer the light into the science fiber, again making use of the A\&G camera where a fraction of the calibration light can be seen and used for guiding it to the correct pixel position.

The optical parameters are summarized in Table\,\ref{tab:FE_Req}:
\begin{table}[h]
\centering
\begin{tabular}{|c|c|}
\hline
\textbf{Parameter}                                      & \textbf{Value}            \\ \hline
Telescope focal length                                  & 8400 mm                   \\ \hline
Telescope Entrance Pupil Diameter                       & 1200 mm                   \\ \hline
Waveband                                                & 380 – 680 nm              \\ \hline
Focal ratio on fiber                                    & $f/3.9$                   \\ \hline
fiber diameter                                          & 50 $\mu$m                 \\ \hline
Atmospheric Dispersion Corrector                        & Required                  \\ \hline
Collimator and A\&G field on sky                        & 400 arcsec square         \\ \hline
Collimator and A\&G field at telescope                  & 16.287 mm square          \\ \hline
A\&G camera                                             & $2048 \times 2048 \times 6.5 \mu$m square pixels \\ \hline
Collimator and A\&G field at fiber                      & 9.074 mm square           \\ \hline

\end{tabular}
\caption{Spectroscopy mode optic parameters}
\label{tab:FE_Req}
\end{table}

\subsection{Optical Implementation}
The collimator is based on a Petzval construction of two positive groups using OHARA glasses with a focal length of 200 mm. The first group is a doublet of BSL7Y (the “flint”) and S-FPL53 (the “crown”) while the second group is a triplet of the same glasses plus a weak element of PBM2Y (strong “flint”). This glass combination allows for an excellent state of chromatic correction\footnote{http://www.primeoptics.com.au/Reference/dRPD\_description.pdf}. The ADC doublet prisms consist of 2 OHARA glasses, S-FPL51Y and PBM2Y. They are quite different in base index, so some tilt (0.15\textdegree) of the external surfaces is necessary to maintain a stationary image as the zenith angle changes. The CRM follows the ADC. This mirror directs light from a calibration fiber in the fiber block back to the science fiber in the center of the block. The fiber array fits inside a circle of diameter 6 mm, well within the collimator field of view on the fiber imager image surface. The TTM follows the CRM to allow for easier packaging of the entire assembly.

The beamsplitter cube is positioned so as to redirect some light from the star to the A\&G camera and simultaneously some light from the guide fibers to a retro mirror, and thence to the A\&G camera. The A\&G camera and the fiber imager are also based on the same Petzval construction as the collimator, with focal lengths of 163.6 mm and 111.4 mm respectively.

\subsection{Performance}
\begin{figure}[!h]
    \centering
    \includegraphics[width = 0.49 \textwidth]{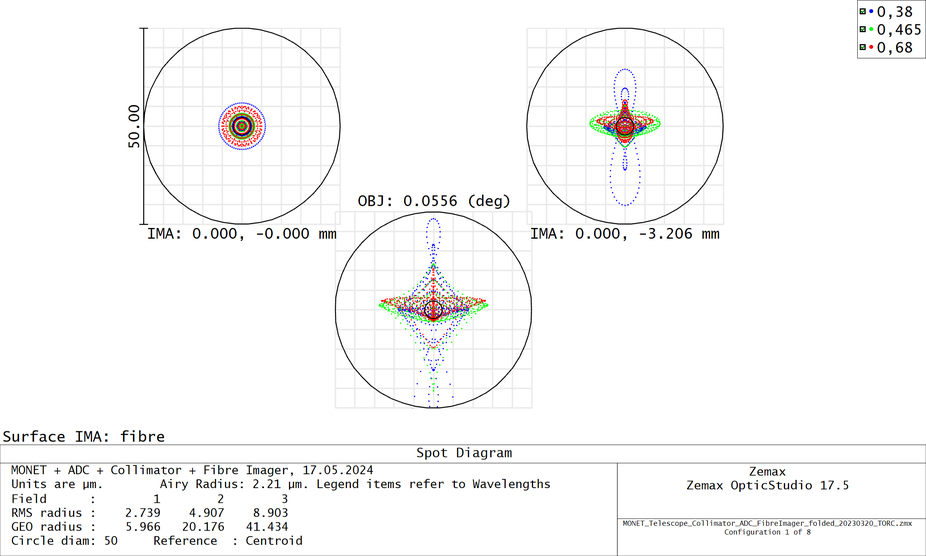}
    \includegraphics[width = 0.41 \textwidth]{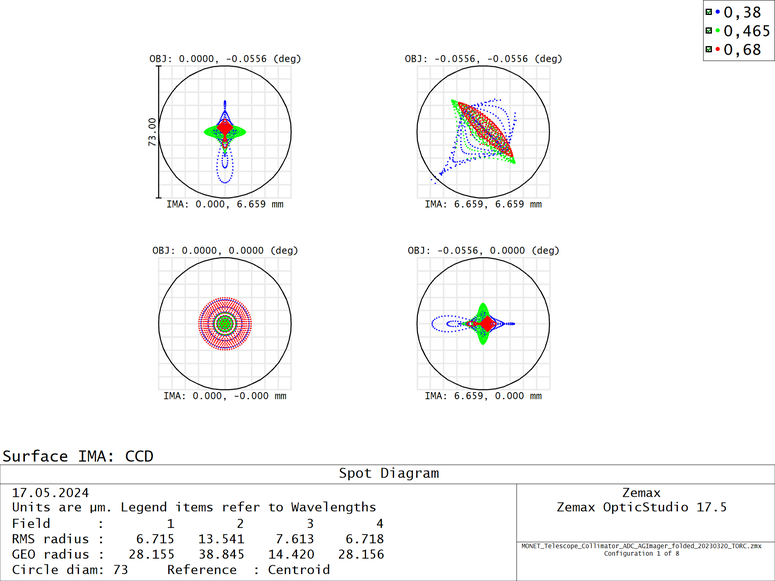}
    \vspace{0.2cm}
    \caption{Left: Optical performance of the science mode for the center, 141 and 200\,arcsec off center positions. Right: Optical performance of the guiding for the 400\,arcsec field of view. }
    \label{fig:FE_perform1}
\end{figure}

Spot diagrams in science and calibration mode are shown in Figure\,\ref{fig:FE_perform1}. They illustrate the performance of the imagery on the 50\,$\mu$m fiber, which shows a RMS diameter of around 10 \% of the fiber diameter over the range of zenith angles.

\subsection{Optomechanical Layout}
\begin{figure}[!h]
    \centering
    \includegraphics[width = 0.42 \textwidth]{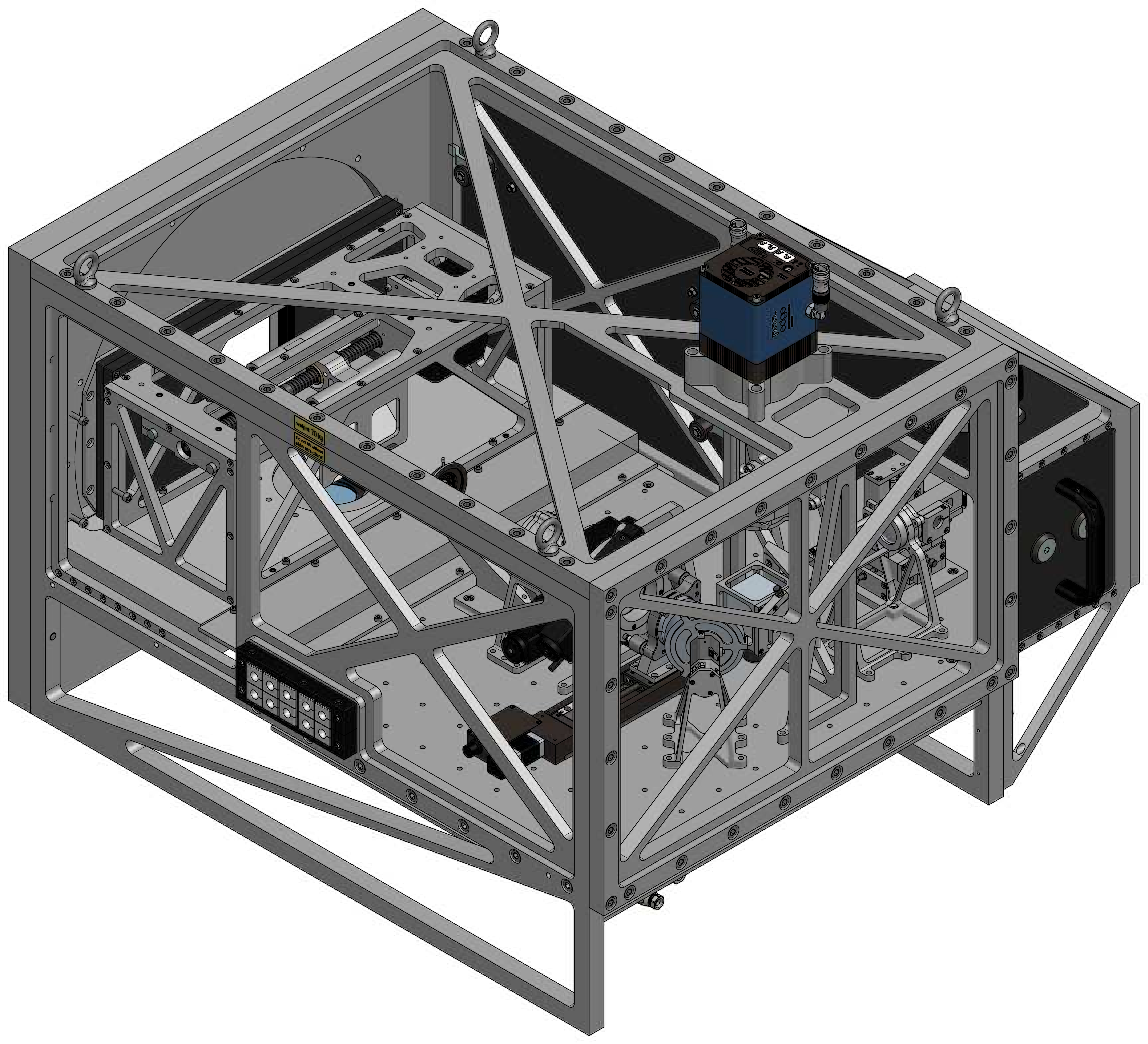}
    \input{pictures/FE-Optics-labeled}

  \vspace{0.2cm}
    \caption{CAD image of the FE unit without the cover plates.}
    \label{fig:FE_cad}
\end{figure}

The new FE (see Fig.\,\ref{fig:FE_cad}) is directly attached to the 340\,mm flange of the de-rotator located at one of the Nasmyth ports of the telescope. It is designed to be as lightweight and as rigid as possible to avoid flexures that could occur when the de-rotator turns the FE unit. The main structure is made from black anodized aluminum, while the light-tight cover plates (not shown in Fig.\,\ref{fig:FE_cad}) are made of thermal insulating Alu DIBOND sandwich material (aluminum covers with a PE core). 

\begin{figure}[!h]
    \centering
    \includegraphics[width = 1.0 \textwidth]{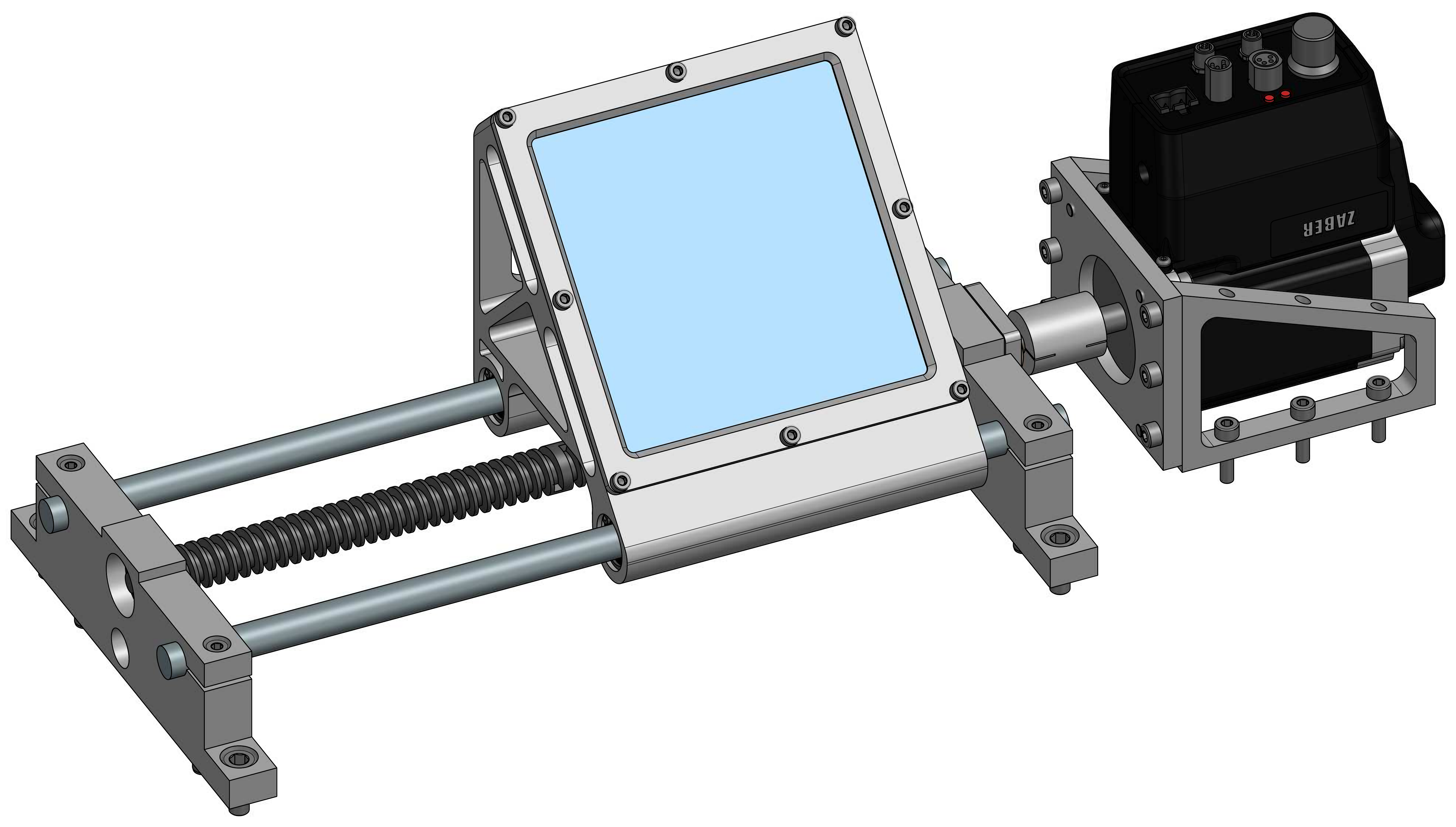}
    \caption{CAD image of the instrument selction unit.}
    \label{fig:FE_select}
\end{figure}

The instrument selection mirror (Fig.\,\ref{fig:FE_select}) is moved by a ZABER X-NMS23-E Stepper Motor motor, the ADC prisms are mounted inside modified Standa 8MR190-2 motorized rotation stages, and the fast tip tilt mirror is sitting on a npoint N-RXY6-SS-521 tip tilt stage. Its closed loop provides a stroke of $\pm6$\,mrad which is enough for all fast corrections needed for optimal fiber injection. While in theory capable of 50\,Hz movement the weight of the mirror will limit this to a lower number, which will be tested before installation. The guiding camera, a pco.edge\,4.2\,BI, is fast enough to feed the loop up to 40\,fps. In combination this setup will be fast enough to compensate for all remaining flexures inside the FE and also correct for guiding errors of the telescope.

Both the guiding camera and the CMOS for the photometric mode are mounted on the outside of the FE for easy access, straightforward electrical wiring, and to keep waste heat outside the FE. All cables from the inside of the FE unit (including the 4 single mode fibers, the calibration fiber and the science fiber) will be fed to the wall using Icotek cable feedthroughs, making the whole unit light and dust protected.

\clearpage
\section{Spectrograph}
\label{sec:spectrograph}
\subsection{Overview}
\begin{table}[h]
\centering
\begin{tabular}{|l|c|}
\hline
\textbf{Parameter} &  \\ \hline

Input science fiber & $50 \mu$m octagonal @ f/3.6 \\ \hline
Spectrograph slit after pupil slicer and double scrambler & $33\times132 \mu$m  @f/4.4 \\ \hline
Virtual slit after f/n conversion & $75 \mu$m $\times$ 0.6 mm @ f/10 \\ \hline
Waveband & 380 – 680 nm \\ \hline
Sampling at center wavelength, 1 resolution element & 3.5 px \\ \hline
Hi res echelle  & RGL MR263 replica \\ \hline
 Resolving power R & $>$ 82000 \\ \hline
\end{tabular}
\caption{Spectrograph parameters}
\label{tab:spec}
\end{table}

\subsection{Optical design}

The spectrograph is based on a triple-pass, reflective, white pupil echelle spectrograph of focal length of 0.825\,m and focal ratio of 10. A 0.05\,mm diameter octagonal fiber captures a 2.2\,arcsecond field at the telescope focal plane and transports the stellar light into the spectrograph. Here a combined pupil slicer and double scrambler (see \cite{Beckert2022}) reshapes the beam into a rectangular fiber (33x132\,µm) that forms the slit of the spectrograph. An additional quadratic 33x33\,µm calibration fiber will come directly from the calibration unit (see Section\,\ref{sec:CU}) and is placed on top of this slit. A fiber-slit relay will convert the 33\,µm wide fiber's f/4.4 output into a 75\,µm wide virtual slit at f/10 for the spectrograph.

\begin{center}
\begin{figure}[h!]
\begin{center}
\begin{tikzpicture}[scale=1.176]

  \node[anchor=center] at (0,0) {\includegraphics[width = 1 \textwidth]{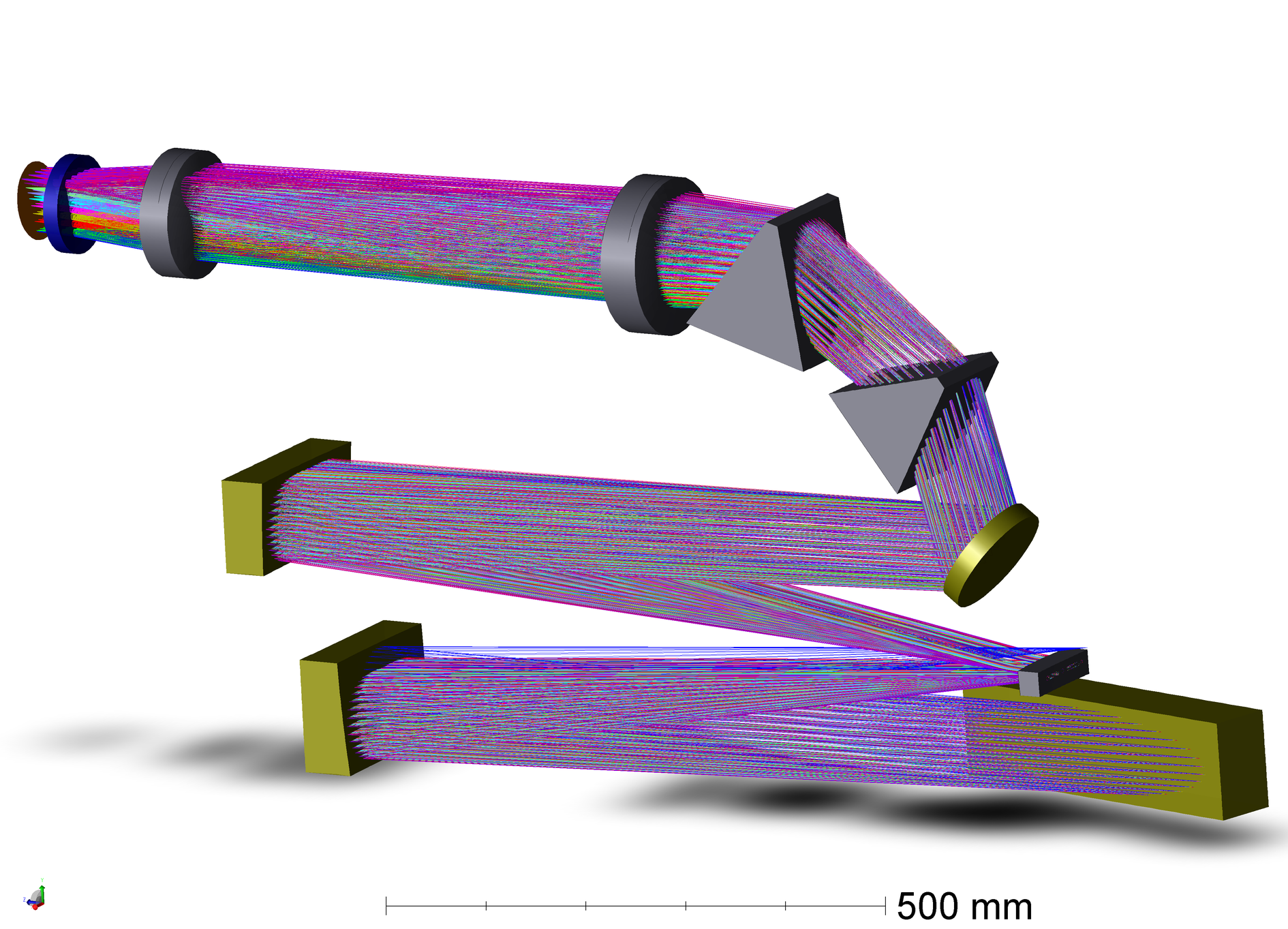}};

  \node[Labels](A1) at (-4,4.5){\textbf{Camera}};
    \draw[ARROW] (A1)--++(270:0.5cm)--(-0.15,3.3cm);
    \draw[ARROW] (A1)--++(270:0.5cm)--(-4.8cm,3.5cm);
    \draw[ARROW] (A1)--++(270:0.5cm)--(-6.2cm,3.5cm);
  
  \node[Labels](A2) at (-4.5,1.3){\textbf{Detector}};
    \draw[ARROW] (A2)--++(180:1.5cm)--(-6.95cm,2.5cm);
    
  \node[Labels](A3) at (0,0.75){\textbf{White pupil}};
    \draw[ARROW] (A3)--++(0:1.8cm)--(2.2,1);
  
  \node[LabelsLong](A4) at (4,4){\textbf{2 cross disperser prisms}};
    \draw[ARROW] (A4)--++(270:0.75cm)--(2.3,3);
    \draw[ARROW] (A4)--++(270:0.75cm)--(4,1.3);
    
  \node[Labels](A5) at (5.5,1){\textbf{Fold mirror}};
    \draw[ARROW] (A5)--++(270:0.65cm)--(4.5,-0.5);
    
  \node[Labels](A6) at (6,-0.5){\textbf{Virtual slit}};
    \draw[ARROW] (A6)--++(270:0.55cm)--(5,-2);
    
  \node[Labels](A7) at (3.5,-4.3){\textbf{Echelle}};
    \draw[ARROW] (A7)--++(0:1.5cm)--(5.5,-4);
    
  \node[Labels](A8) at (6.3,-2){\textbf{CFF}};
    \draw[ARROW] (A8)--++(180:0.9cm)--(5.12,-2.2);
    
  \node[LabelsLong](A9) at (-5,-4.2){\textbf{Triple pass collimator mirror}};
    \draw[ARROW] (A9)--++(90:0.8cm)--(-4.5,-1.3);
    \draw[ARROW] (A9)--++(90:0.8cm)--(-4,-3);


  

\end{tikzpicture}

    \caption{Spectrograph configuration.}
    \label{fig:spec1}

\end{center}
\end{figure}
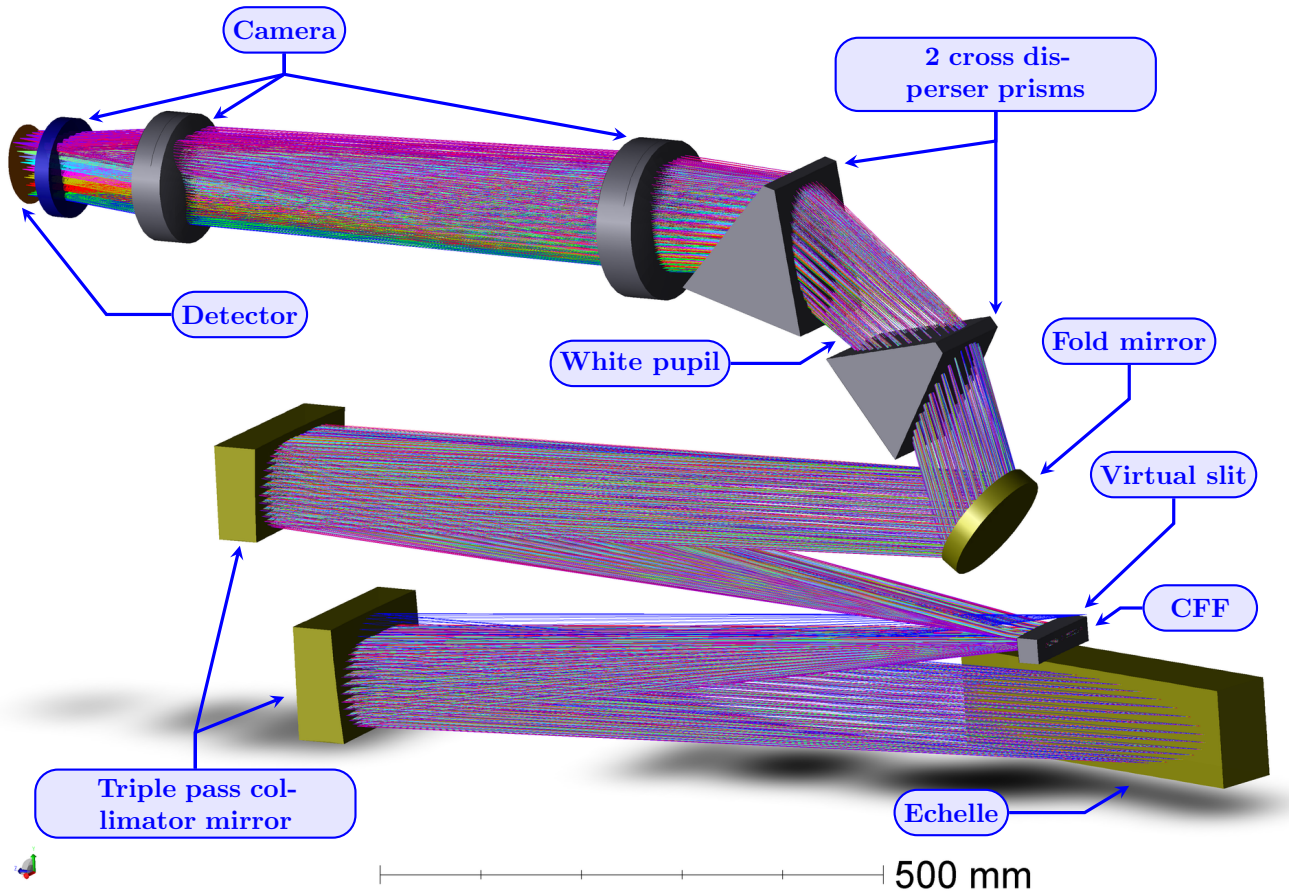    
\end{center}

Figure\,\ref{fig:spec1} illustrates the spectrograph configuration starting at the rectangular slit at f/10. The spectrograph collimator is one half of a single paraboloidal mirror, about 360,mm in diameter. The first pass of the collimator mirror collimates the light from the slit for the echelle. The second pass of the collimator mirror focuses the dispersed light to a real focus. A catadioptric field flattener (CFF) element is placed ahead of this focus to minimize the intrinsic field curvature and field aberrations of the triple pass spectrograph and preserve the pupil imagery. The CFF is a fused silica (for example, Corning HPFS) positive meniscus lens, convex to the collimator mirrors, with a reflective rear face. It is placed ahead of the focus so as to mitigate any possible issues with inhomogeneities in its bulk that could cause troublesome beam distortions of the small beam footprints. The CFF reduces the intrinsic field aberrations and field curvature of the triple-pass paraboloidal spectrograph to negligible levels, which is essential for high-resolution spectroscopy. The third pass, using the other half of the collimator mirror, recollimates the dispersed light and forms the real spectrograph exit pupil (the “white pupil”). This half of the paraboloidal collimater mirror is displaced axially from its twin because the CFF is placed in front of the first focus, breaking the symmetry. 

Two prisms of a high dispersion i-line glass (OHARA PBM2Y) are placed near the white pupil. These are required to provide the necessary cross-dispersion and clearance between the monochromatic fiber images at the long wavelength orders on the image surface. They are tipped a little away from minimum deviation so as to provide some anamorphism, which compresses the slit images in the cross-dispersion direction so that all orders can fit on the detector. 

A f/7 camera images the cross-dispersed spectra and is based on a field-flattened Petzval construction with three components. The camera is composed of OHARA i-line (PBM2Y, BAL15Y) or high transmission glasses (S-FPL51, fused Silica\footnote{http://www.primeoptics.com.au/Reference/dRPD\_description.pdf}). All surfaces are either spherical or plane. The required image quality (IQ) in the high dispersion direction is specified as an 80\% encircled energy width of around one-quarter of a resolution element - approximately 13,µm. This translates to a 1D RMS half-width of approximately 5\,µm. The required IQ in the cross-dispersion direction is at least 50\% greater and is usually not a consideration.

\begin{figure}
    \centering
    \includegraphics[width = 1 \textwidth]{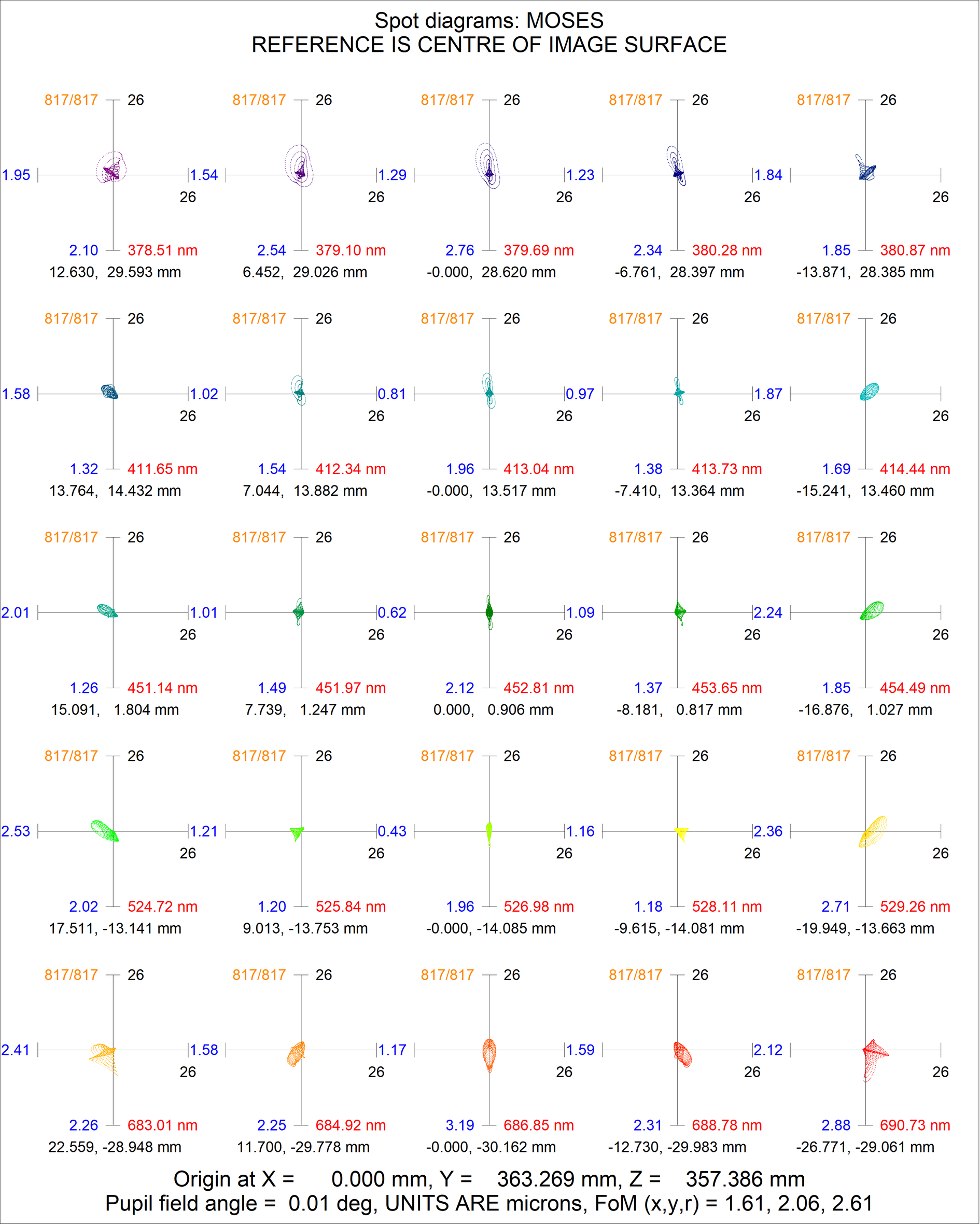}
    \caption{Spectrograph performance. Orders from top to bottom are: 161, 148, 135, 116, 89. Rows span 1.0 FSR, plot dimensions are 1 resolution element.}
    \label{fig:spec2}
\end{figure}
 
It can be seen from Figure\,\ref{fig:spec2} that the average IQ is well within the requirement, leaving plenty of room for a generous error budget. It is also anticipated that the ensemble will be assembled in a deterministic way; thereby considerably easing pressure on the error budget. 

\subsection{Ghost and Stray Light Considerations}
A comprehensive ghost analysis did detect some interesting features of the optical configuration. There are a couple of camera component ghosts of diameter 0.4\,mm or greater which were not deemed to be problematic. These would be mitigated via good AR coatings on the lenses. In this system, CCD ghosts are a potential problem when the ghost light can propagate back to the echelle and be recombined to form "white" ghosts back on the CCD. In this case, the tilt of the CCD, and the camera and prism apertures means that only a small portion of a couple of the short wavelength orders will make it back to the echelle. The recombination ghost so formed is then extinguished completely at the CFF aperture. CFF ghosts can also be problematic. These are effectively extinguished by careful choice of surface tilts (wedge) in the CFF so that the ghost light is directed onto baffles outside the collimator mirror apertures.

Diffraction gratings are not hugely efficient devices and there will be significant spillage into the collimator cavity. Secondary scattering will be damped by the use of textured baffles (for grazing incidence) and good "blacks" on surfaces that could be exposed to the grating spillage. It goes without saying that all components, and especially the mirrors, will have a high quality polish.

\subsection{CCD detector system}
The detector will be a 4kx4k STA4850 CCD and is accompanied by a detector system including the cryostat, the Archon CCD controller, and a CryoTel CT Stirling cryocooler by SunPower. The STA4850, with its 15\,$\mu$m pixel size, is the same that is used in the MAROON-X instrument where it has shown excellent performance in this exact configuration (see \cite{seifahrt2018maroonx}).

\section{Calibration}
\label{sec:CU}
The calibration scheme is adapted from precious work on CARMENES (see for example \cite{Bauer2015}, \cite{Schafer2018-cq} and \cite{Bauer2020-oc}) and involves hollow cathode lamps (HCL) for determining the wavelength solution, a flatfield lamp and a Fabry-Pérot etalon system (FP) for both drift corrections of the spectrograph and improving the wavelength solution where the HCL has a lower line density. 

Daily calibration sequences at dusk and dawn will consist of HCL in the science fiber (send through the FE, see Sec.\,\ref{sec:frontend} and the FP in the calibration fiber. During the night the science target will be in the science fiber and the FP will always be observed simultaneously in the calibration fiber. The FP unit itself will be another iteration of the newest CARMENES FP unit. This setup has already shown a RV precision performance of better than 10\,cm/s over 40\,h \citep{Debus2024}.

\bibliography{spie2024} 
\bibliographystyle{plainnat} 

\end{document}